\pdfoutput=1

\documentclass[conference]{IEEEtran}
\usepackage{cite}

\ifCLASSINFOpdf
  \usepackage[pdftex]{graphicx}
  \graphicspath{{img/}}
  \DeclareGraphicsExtensions{.pdf,.jpeg,.png,.PNG}
\else
  \usepackage[dvips]{graphicx}
  \graphicspath{{eps/}}
  \DeclareGraphicsExtensions{.eps}
\fi
\usepackage{amsmath}
\usepackage{algorithmic}

\usepackage{array}

\usepackage[caption=false,font=footnotesize]{subfig}
\usepackage{fixltx2e}
\usepackage{dblfloatfix}

\usepackage{url}

\usepackage{algorithm,algorithmic}
\usepackage{comment}
\usepackage{booktabs}
\usepackage[english]{babel}

\usepackage{fancyhdr}
\begin{document}
%
\title{Unidirectional Quorum-based Cycle Planning for Efficient Resource Utilization and Fault-Tolerance \\ \footnotesize The final publication is available at IEEE via \url{http://dx.doi.org/10.1109/ICCCN.2016.7568595}}

\author{\IEEEauthorblockN{Cory J. Kleinheksel, Member, IEEE and Arun K. Somani, Fellow, IEEE}
\IEEEauthorblockA{Electrical and Computer Engineering, Iowa State University, Ames, Iowa 50011\\
Telephone: (515) 294--0442, Fax: (515) 294--9273, Email: \{cklein, arun\}@iastate.edu}}
\maketitle

\begin{abstract}
In this paper, we propose a greedy cycle direction heuristic to improve the generalized $\mathbf{R}$ redundancy quorum cycle technique.  When applied using only single cycles rather than the standard paired cycles, the generalized $\mathbf{R}$ redundancy technique has been shown to almost halve the necessary light-trail resources in the network.  Our greedy heuristic improves this cycle-based routing technique's fault-tolerance and dependability.

For efficiency and distributed control, it is common in distributed systems and algorithms to group nodes into intersecting sets referred to as quorum sets.  Optimal communication quorum sets forming optical cycles based on light-trails have been shown to flexibly and efficiently route both point-to-point and multipoint-to-multipoint traffic requests. Commonly cycle routing techniques will use pairs of cycles to achieve both routing and fault-tolerance, which uses substantial resources and creates the potential for underutilization.  Instead, we use a single cycle and intentionally utilize $\mathbf{R}$ redundancy within the quorum cycles such that every point-to-point communication pairs occur in at least $\mathbf{R}$ cycles.  Without the paired cycles the direction of the quorum cycles becomes critical to the fault tolerance performance.  For this we developed a greedy cycle direction heuristic and our single fault network simulations show a reduction of missing pairs by greater than 30\%, which translates to significant improvements in fault coverage.
\end{abstract}


%
\IEEEpeerreviewmaketitle

\section{Introduction}

Internet backbones and data centers have become key pieces of national infrastructure rising almost to the level of roads, bridges, electricity, and water.  Optical networks in these settings are depended upon for high speed communications. 
Failures within a network are to be expected and can happen as much as every couple days\cite{dlastine2012fault}.  

Knowing the unicast or multicast requests a priori is often not possible.  This constraint makes protection against faults in those arbitrary communication paths a challenge.  An efficient all node pairs protection scheme supporting both unicast and multicast communication is necessary.

For efficiency and distributed control, it is common in distributed systems and algorithms to group nodes into intersecting sets referred to as quorum sets.  Quorum-based cycle routing can efficiently support arbitrary point-to-point and multi-point optical communication\cite{dlastine2014thesis}.  Cycles are created using quorums of nodes. Within a cycle, multicasts to all nodes in that cycle are possible. The proof that cyclic quorums have an ``all-pairs'' property guarantees unicast capabilities in a network using the quorum cycles routing technique\cite{ckleinheksel2016scaling}.  Efficient broadcasts can be achieved with $O(\sqrt{N})$ multicasts by exploiting the same properties.

Fault tolerance of optical networks with quorum-based cycle routing has been analyzed\cite{ckleinheksel2015quorum}.  Observing communication pairs were being protected by more than one cycle, $R$ redundancy quorums were used to improve resource efficiency\cite{ckleinheksel2015Enhancing}.  Resource efficiency can come at a cost and in this research, we propose a heuristic to improve the redundant quorum cycle routing techniques.  Our single fault network simulations show a reduction of missing pairs by 31.16 - 48.74\% and 35.48 - 44.85\%, $R=2$ and $R=3$ respectively, which translates to significant improvements in fault coverage.

The next sections describe the network model, light-trails, and cycle routing.  Section \ref{sec:Quorums:sub:Routing} establishes how quorums can be used for optical network routing. We analyze the current strengths and opportunities for improvements to $R$ redundant quorum-based single cycle routing in Section \ref{sec:singleCycleAnalysis}.  Lastly in Section \ref{sec:improvingHeuristic}, we present a greedy cycle direction heuristic to improve the optical network fault tolerance.

\section{Network Model} \label{sec:commBackground:networkModel}
\label{sec:NetworkModel}


The fiber-optic networks consist of several transmitters and receivers (nodes) interconnected by fiber-optic cables.  The cables form the links (i.e., edges) between those nodes, which leads to a convenient model of a network in terms of a graph $G = (V,E)$.  $V$ are the set of nodes in the network and $E$ are the set of edges. 

Edge $(e_i,e_j)$ is a fiber-optic link connecting nodes $e_i$ and $e_j$ in the network, where $e_i,e_j \in V$ and $(e_i,e_j) \in E$.  It is a general assumption that the same set of optical wavelengths are available on all edges in $E$.  The number of wavelengths available per optical fiber is dependent on the fiber-optic cables and the transmitter/receiver pairs.

\subsection{Light-Trails} 
\label{sec:light_trailDefinition}

Lightpaths were a critical building block in the first optical communications, but required significant traffic engineering and aggregation to support point-to-point communication, or pay the penalty of low resource utilization on the fiber-optic link.  Light-trails were proposed \cite{chlamtac2003light} as a solution to the challenges facing lightpaths and could be built using commercial off-the-shelf technology. Significant contributions have been made to enable adoption and advance the light-trail architecture\cite{fang2004optimal,li2008multicast,dlastine2011ECBRA}.  Light-trail communication is all optical and uses the same wavelength(s) from start to end node; avoiding Optical-to-Electrical-to-Optical (O/E/O) conversions, energy inefficiencies and time delays at intermediate hops.

Light-trails enable fast, dynamic creation of a unidirectional optical sharable communication channel.  This communication channel allows for channel receive and transmit access to all connected nodes, making them more suitable for IP-centric traffic \cite{fang2004optimal}.  Communications from an upstream node to one or more downstream nodes can be scheduled.  The scheduling protocol avoids collisions within a light-trail and controls when nodes are able to transmit to downstream nodes.  The scheduling is generally assumed to occur over a control channel, which may or may not be separate from the shared optical fiber that is being used for the light-trail.

Optical shutters allow for wavelength reuse within the network. Start and end nodes have their optical shutters in the \textit{off} state, while intermediate nodes have their optical shutters in the \textit{on} state.  This effectively isolates an optical signal to a specific light-trail and allows for reuse of optical wavelength(s) elsewhere in the network.

Nodes can receive from the incoming signal while the signal is simultaneously continuing to downstream nodes, sometimes referred to as a drop and continue function.  Early technology supported only a few wavelengths; however, the latest devices may support over 100 channels, hence allowing multiple light-trails to share the same edge in the network for a combined over 1-Terabits/s\cite{agrawal2007nonlinear}. 


\section{Light-Trails, Cycle Routing, and Fault Tolerance}
\label{sec:cycleRouting}

\begin{figure}[t]
	\centering
	\includegraphics[width=2.5in]{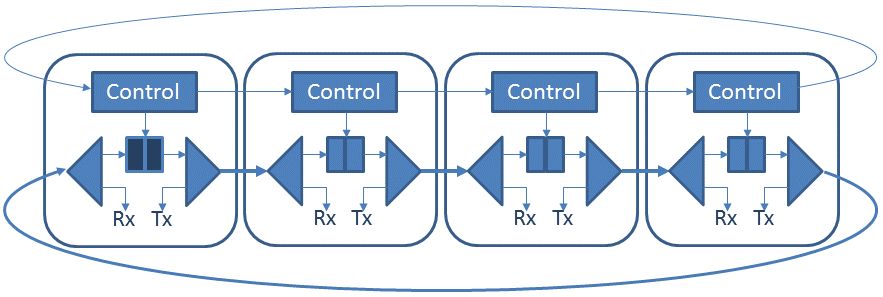}
	\caption{Cycle formed using the light-trail architecture}
	\label{fig:lighttrail_cycle}
\end{figure}

Point-to-point and multi-point traffic requests have a set of nodes $C = \{e_i,...,e_j\}$ that wish to communicate and need to be protected against network faults.  Establishing a primary and backup multicast path from every node to every other node in $C$ can be a waste of resources. 
We utilize the light-trail architecture in the form of a cycle (Fig. \ref{fig:lighttrail_cycle}).  The cycle will both route the multi-point request and protect it at the same time using fewer resources.

Figure \ref{fig:lighttrail_cycle} is a light-trail where the start and end node is the same node, referred to as the \textit{hub node}.  The hub node has its optical shutters in the \textit{off} state, while intermediate nodes have their optical shutters in the \textit{on} state.  




Failures within an optical network are to be expected.  SONET rings can be used to protect point-to-point and shared paths while enabling failure location.  The generalization of p-cycle protection to allow for path and link protection was proposed in \cite{grover2003extending}.  P-cycle protection of unicast and multicast traffic networks requires preconfiguration, and the offline nature allows for the efficient cycles to be selected. 
The Optimized Collapsed Rings (OCR) single link protection heuristic was developed to address the heterogeneous, part multicast / part unicast, nature of WDM traffic\cite{khalil2005pre}.

The multi-point cycle routing algorithm (MCRA) uses bidirectional cycles for fault tolerance and is capable of supporting SONET rings and p-cycles\cite{dlastine2012fault}.  ECBRA is a significant improvement of MCRA and outperforms the OCR heuristic\cite{dlastine2011ECBRA}. ECBRA heuristic balances optimality and speed, taking $O\left(\left|E\right|\left|C\right|^3\right)$ steps to find a close to optimal cycle.  

\section{Quorums Sets for Routing}
\label{sec:Quorums:sub:Routing}

Quorums have been used for efficient point-to-point, multicast, and all-to-all traffic requests in optical networks\cite{dlastine2014thesis}.  This is important because traffic in many optical networks is heterogeneous meaning the routing framework must be able to handle all types.

Quorums sets cover $N$ entities, in this case $N=\left|V\right|$ optical network nodes.  The number of entities, $N$, also defines the number of small subsets, i.e., quorums, that will be used in our solution.  

A quorum set minimally has the property that all quorums must intersect.  Specifically for distributed implementations, it is also desirable that each quorum have equal work and equal responsibility within the quorum set.  Cyclic quorum sets have these properties\cite{luk1997two} and we use them for efficient communication.

Point-to-point, multicast, and all-to-all traffic can be routed through an optical network with $N$ cycles based on cyclic quorums\cite{dlastine2014thesis,ckleinheksel2015quorum}.  Cyclic quorums sets are used as the basis for cycle routing to guarantee all possible node pairs occurred in at least one cycle\cite{ckleinheksel2016scaling}.  When performing a multicast, if participants belong to the same cycle, then one cycle can be used.  Realistically, requests will span multiple quorums and/or be larger than a single quorum cycle.  The worst case becomes a broadcast with an upper bound of requiring no more than $k$ cycles to route and protect broadcast traffic.

\subsection{Defining Cyclic Quorum Sets}\label{sec_define_cyclic_quorum_sets}

Cyclic quorum sets are based on cyclic block design and cyclic difference sets.  However, searching for optimal sets requires an exhaustive search\cite{luk1997two}.  Cyclic quorum sets are unique in that once the first quorum (Eq. \ref{equ_first_quorum}) is defined the remaining quorums in the set can be generated via incrementing the indices (modulus to keep indices within bounds is not shown in Equation \ref{equ_general_quorum} for conciseness).  

\begin{equation}\label{equ_first_quorum}
S_0=\left\lbrace e_0,\dots,e_j \right\rbrace
\end{equation}
\begin{equation}\label{equ_general_quorum}
S_i=\left\lbrace e_{0+i},\dots,e_{j+i} \right\rbrace, \:\forall\: i\in0,1,\dots,N-1
\end{equation}
For simplicity, assume $e_0\in S_0$ without loss of generality (any one-to-one re-mapping of indices can result in this assumption).  

For our work, we used the $N=4,\dots,111$ optimal cyclic quorums from \cite{luk1997two}. 

\subsection{Redundant Cyclic Quorums Sets}
\label{chpt:Quorums:sub:Redundant}

In this section we define $R$ redundant cyclic quorums sets as proposed in \cite{ckleinheksel2015Enhancing}.  Quorum-based cycle routing solutions used cyclic quorums to form a set of communication cycles which were shown to be almost fault tolerant in fiber optic networks\cite{dlastine2014thesis,ckleinheksel2015quorum}.  

There are $N$ quorums in a quorums set and each quorum has $k$ nodes.  In analysis of networking capabilities, we are interested in whether every node can communicate with every other node (all-pairs). Equation \ref{eq:numberPairs} considers the number of pairs within a quorum, e.g., the pairs made between $k$ nodes communicating with $(k-1)$ other nodes in a single quorum.  Equation \ref{eq:totalNumberPairs} considers the total pairs formed by all $N$ quorums in the set.  For convenience we set $M$ to be the total pairs for a given network with $N$ nodes and $k$ optimal quorum size.

\begin{align}
\frac{k(k-1)}{2}&=O(k^2)\label{eq:numberPairs}\\
N\frac{k(k-1)}{2}&=O(Nk^2)\label{eq:totalNumberPairs}\\
M&=O(Nk^2)
\end{align}

When the quorum size, $k$, is minimal or larger, every pair of nodes $(e_i,e_j)$ occurs together within a quorum in the set at least once. Optical networking, however, requires all directional point-to-point pairs to exist, i.e., both pairs $(e_i,e_j)$ and $(e_j,e_i)$.  Previously this had been addressed by pairing each cycle with the same cycle and its direction reversed.  

It was observed in \cite{dlastine2014thesis} that the quorum-based cycle routing solution had some node pairings occurring together in multiple cycles and it was proposed that these could be used for load balancing.  As an alternative to that option, \cite{ckleinheksel2015Enhancing} added a requirement that every pair $(e_i,e_j)$ would occur together within at least $R$ quorums rather than just one.  Exploiting the natural occurrence of redundant pairs is an attempt to eliminate the need for paired cycles, thus moving the redundancy from the paired cycles and putting the redundancy in the quorums.  

The number of quorums in the solution remained the same $N$, hence to create the additional pairs the quorum size had to be enlarged to $\hat{k}$.  Equation \ref{eq:totalR2NumberPairs} calculates the number of node pairs in quorums of size $\hat{k}$.  Equation \ref{eq:RtimesR1} is our requirement that the total number of pairs have increased $R$ times from the original total pairs, $M$.  Finally, Equation \ref{eq:Rsize} solves for size $\hat{k}$ in relation to optimal $k$.

\begin{align}
N\frac{\hat{k}(\hat{k}-1)}{2}&=O(N\hat{k}^2)\label{eq:totalR2NumberPairs}\\
O(N\hat{k}^2)&=RM\label{eq:RtimesR1}\\
\hat{k}&\approx\sqrt{R}k\label{eq:Rsize}
\end{align}

This result is powerful.  The number of node pairs increased by $R$ times, but the size of $k$ only increased by a factor of $\sqrt{R}$.  Using this reduced growth rate to our advantage, many node pairs can be created without substantially increasing the resources used.  Additionally, this growth rate is far slower than simply duplicating a cycle, hence significantly challenging the need for paired cycles and opening the door for considerable resource savings. 

To the best of our knowledge, no efficient algorithm is known to find quorums of minimum size, particularly with the additional requirement that entity pairs appear a minimum $R$ times within the quorums set solution.  The authors in \cite{luk1997two} used a brute force search to find optimal cyclic quorums for $N=4\dots111$.  Using our generalized result from Eq. \ref{eq:Rsize}, we too used a brute force search beginning with the smallest possible quorum size for a given number of nodes $N$ and a given desired redundancy factor $R$.

The resulting redundant quorums were utilized in following section as we analyzed the efficiency and fault tolerance of quorum-based cycle routing in optical networking.

\section{Redundant Cyclic Quorums Set - Single Cycle Network Analysis}
\label{sec:singleCycleAnalysis}

Optical networks are highly depended upon.  The fault tolerance aspect of these route designs are important.  In \cite{dlastine2014thesis,ckleinheksel2015quorum}, ECBRA was used to route each of the quorums-based cycles.  It was shown that the quorums set approach provided fault tolerance and \cite{dlastine2014thesis} showed that this technique required far fewer links to accomplish the routing of all-to-all traffic, when compared to using point-to-point connections.  

As an unintended benefit, some quorums sets resulted in node pairs occurring in more than one quorums-based cycle.  It was these occurrences of node pairs multiple times that improved the fault tolerance performance.  

Redundant node pairs can also be generated intentionally as described in Section \ref{chpt:Quorums:sub:Redundant}. These can be used to eliminate the paired cycle implementations and use only a single cycle, significantly reducing the required amount of network resources while still maintaining a similar level of fault-tolerance.  Our work reported in\cite{ckleinheksel2015Enhancing} examined this approach.

In this section, we analyze the strengths to deliver resource efficient routing solutions, while identifying opportunities to improve fault tolerance in the network.

\subsection{Fault Model}
\label{sec:FaultTolerance:sub:model}

The fault model assumed for our work is the link (edge) failure.  While a simple model, it does cover most real single fault scenarios.  

The most direct fault to consider is the optical link fault.  This occurs when a link is broken, like planned maintenance or the accidental severing during land excavation.  Modeling link faults as a single edge failure is straightforward.

Each modeled node needs a pair of transmitters and receivers for each occurrence in a cycle.  These pairs of devices can fail too.  Short of a natural disaster, pairs will likely fail independently of one another.  When a transmitter/receiver pair fails within a modeled node, the affect on the global network is similar to that link failing.  Modeling as a single edge failure, while not an exact fault mapping, is an appropriate abstraction.

\subsection{Experimental Setup}

\begin{figure*}[!t] 
	\centering
	\subfloat[]{\includegraphics[width=1.5in]{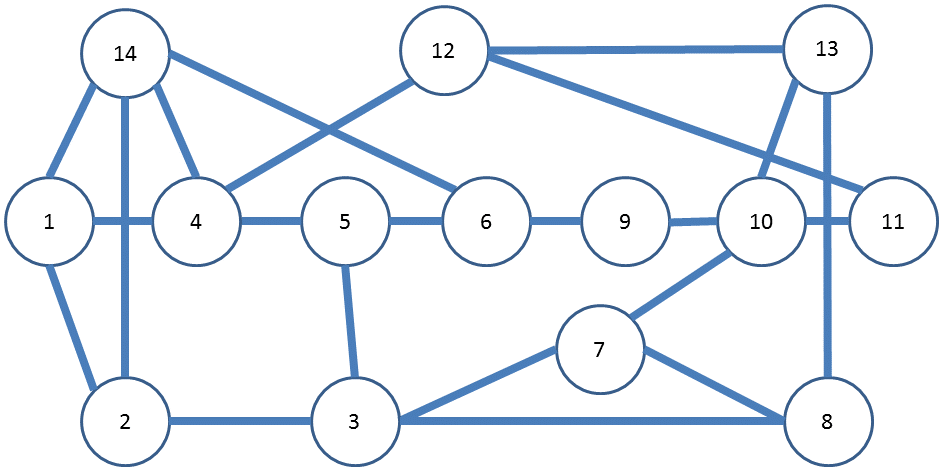}%
		\label{fig:Networks:sub:nsfnet}}
	\hfil
	\subfloat[]{\includegraphics[width=1.5in]{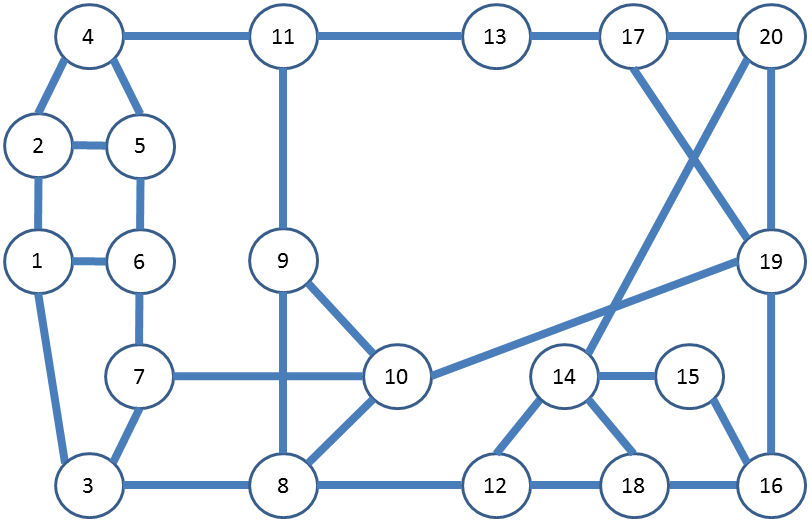}%
		\label{fig:Networks:sub:arpanet}}
	\hfil
	\subfloat[]{\includegraphics[width=1.5in]{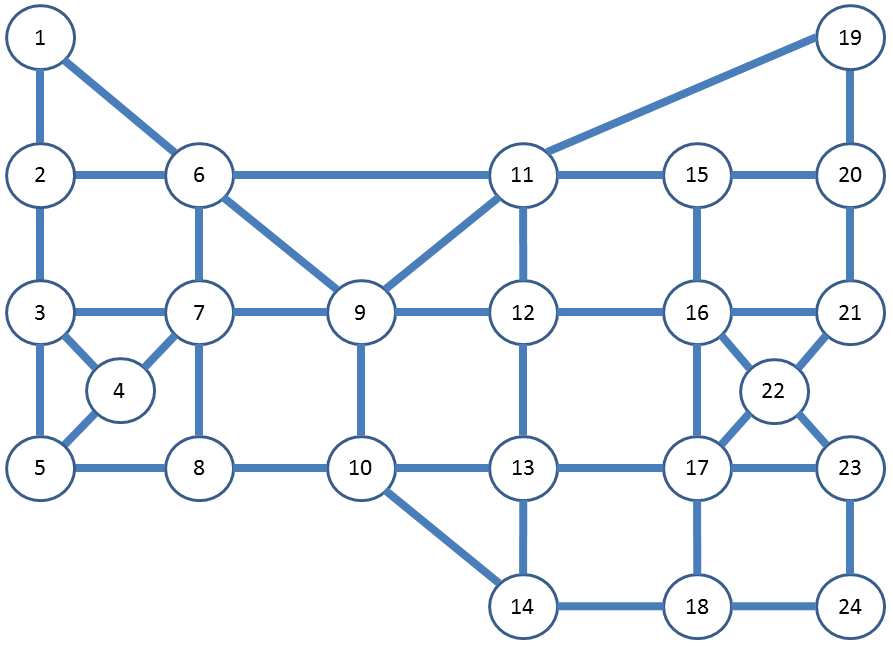}%
		\label{fig:Networks:sub:american}}
	\hfil
	\subfloat[]{\includegraphics[width=1.5in]{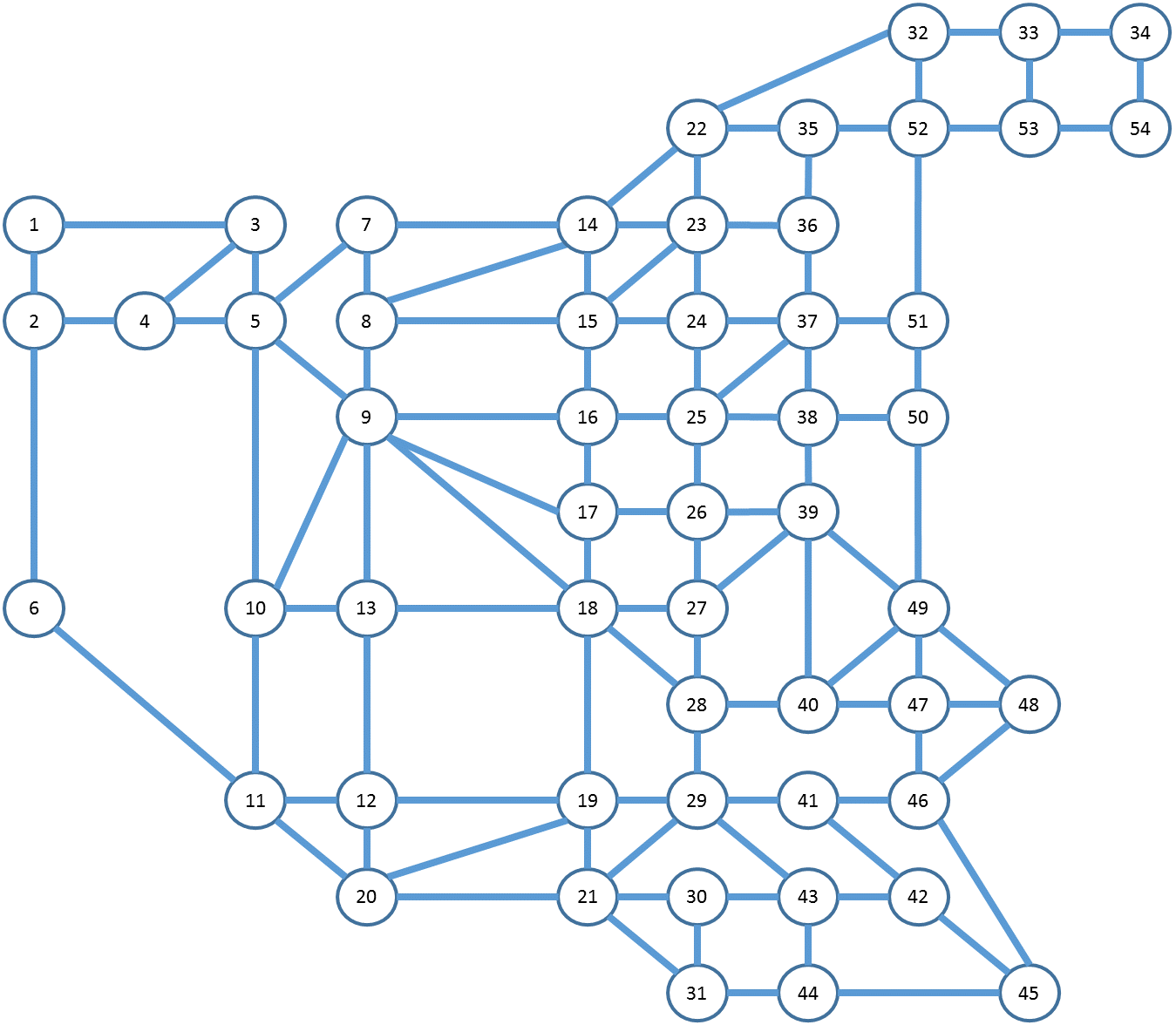}%
		\label{fig:Networks:sub:chinese}}
	\caption{Networks used for simulations: Figure (a) NSFNET, 14-Node/22-Link, Figure (b) ARPANET, 20-Node/31-Link, Figure (c) American Backbone\cite{tang2011multicast}, 24-Node/43-Link, and Figure (d) Chinese Backbone\cite{tang2011multicast}, 54-Node/103-Link.}
	\label{fig:Networks}
\end{figure*} 

\begin{table*}[!t]
	\centering
	\caption{Mean links used by single cycles compared to paired cycles using redundant cyclic quorums (95\% CI)}
	\label{tbl:single:links}
	\begin{tabular}{@{}lrrcrc@{}}
		\toprule
		& 
		$\mathbf{R=1}$ \textbf{(Paired)}    & 
		\multicolumn{2}{c}{$\mathbf{R=2}$ \textbf{(Single)}} & 
		\multicolumn{2}{c}{$\mathbf{R=3}$ \textbf{(Single)}} \\ \cline{2-6} 
		
		\textbf{Network} & 
		\multicolumn{1}{c}{\textbf{Links}}             &
		\multicolumn{1}{c}{\textbf{Links}}             &
		\multicolumn{1}{c}{\textbf{Reduction (\%)}}    & 
		\multicolumn{1}{c}{\textbf{Links}}             &
		\multicolumn{1}{c}{\textbf{Reduction (\%)}}    \\
		\midrule
		
		NSFNET   & 249.32 $\pm$ 1.37  & 135.41 $\pm$ 0.60  & -45.69 & 145.05 $\pm$ 0.74  & -41.82 \\
		ARPANET  & 511.90 $\pm$ 1.87  & 269.63 $\pm$ 0.76  & -47.33 & 294.75 $\pm$ 0.96  & -42.42 \\
		American & 641.38 $\pm$ 2.10  & 360.03 $\pm$ 0.98  & -43.87 & 376.86 $\pm$ 0.77  & -41.24 \\
		Chinese  & 2673.30 $\pm$ 7.11 & 1527.32 $\pm$ 3.50 & -42.87 & 1635.70 $\pm$ 2.99 & -38.81  \\ \bottomrule
	\end{tabular}
\end{table*}

Maintaining the ability to serve all dynamic point-to-point traffic requests despite fault is important.  We examined the fault tolerance of the NSFNET, ARPANET, American backbone, Chinese backbone networks (Fig. \ref{fig:Networks}). 
					
To model the fault, we simulate the failure of each edge, $(e_i,e_j) \in E$, in the network model, $G=(V,E)$.  We then examine the network's ability to serve all potential point-to-point requests by counting pairs of nodes that would be able to communicate and conversely those pairs that are unable to communicate.

\subsection{Network Routing Efficiency and Fault Tolerance}

We examine redundant quorums by using additional cyclic quorum redundancy with just a single cycle compared to paired cycles used in prior art.  The additional redundancy is used to distribute node communication pairs across different cycles, e.g., $0\rightarrow 1$ and $0\leftarrow 1$ may not occur within the same cycle.  Recall that optical light-trail cycles are unidirectional, so prior art would satisfied both $0\rightarrow 1$ and $0\leftarrow 1$ with a pair of cycles.  One cycle would form the forward communication $0\rightarrow 1$, while the second cycle would form the backward communication $0\leftarrow 1$.  By distributing the communication pair responsibility throughout the network, we may only need a single cycle, thus significantly reducing the required amount of network resources.

We used four common networks (Fig. \ref{fig:Networks}) and an implementation of the ECBRA heuristic\cite{dlastine2011ECBRA} to perform the cycle routing.  ECBRA is sensitive to node and edge numbering that a total of 100 variations on the inputs were considered, each being a one-to-one mapping with the respective network.  For simulation of prior art, we used the $N=4,\dots,111$ optimal cyclic quorums from \cite{luk1997two}.  Redundant cyclic quorums for $R=2$ and $R=3$ were found using the techniques described in Section \ref{chpt:Quorums:sub:Redundant} and in our work reported in\cite{ckleinheksel2015Enhancing}.

\subsubsection{Fault-free operational analysis} \label{chpt:COMM:sub:RedundantSingle:Fault-Free}

It is expected that a majority of the time the optical network will be operating without faults.  It is important that the resource utilization during this period be analyzed.

The metric we use to measure resource utilization is the number of links used in a solution.  Comparing network-to-network is not particularly insightful, but comparing multiple solutions for a particular network is.  The more links that a set of quorum cycles use, the fewer (wavelength) resources that can be assigned to each link.  Additionally, each logical link represents a required physical transmitter and receiver, hence capital costs.

Table \ref{tbl:single:links} shows significant 38.81 - 42.42\% resource reduction when using $R=3$ redundancy in quorums over the more traditional, prior art methods using paired cycles.  $R=2$ gives even better resource reduction.  This reduction represents the potential for lower capital costs in terms of physical transmitters and receivers needed and the potential for more (wavelength) resource availability within the network.  The paired cycles results with a 95\% confidence interval (CI) for $R=1$ is in column two of Table \ref{tbl:single:links} for comparison to the single cycle, increased quorum redundancy technique.  The technique uses far fewer links (shown in columns 3 - 6.)  $R=2$ comes close to halving the necessary resources.

Previously paired light-trails were used to form all of the point-to-point communication node pairs. 
We analyze the impact of increasing the $R$ redundancy within quorums and its impact of keeping resource utilization low using the missing node pairs metric.  

Ideally, like the paired cycle case, there would be 0\% missing, however single cycles do not have the benefit of both $(e_i,e_j)$ and $(e_j,e_i)$ pairs occurring in the same cycle.  The dramatic reduction in resource utilization came at a trade off of a few missing communication pairs.  $R=2$ missed 0.95\% or fewer on average (95\% CI), and $R=3$ missed even fewer at 0.21\% or less on average (95\% CI).  Redundant $R=3$ cycles performs approximately 2+ times better every time.  As seen in Table \ref{tbl:single:links}, this performance improvement came at a slightly higher cost, while still being significantly smaller than the state of art approach.

		

It is the limitations of unidirectional optical light-trail cycles with its required one optical shutter in the \textit{off} state per cycle that has caused the missing pairs and the potential need for additional compensation steps.  Compensation is possible using an off-the-shelf solution of an additional routing step involving an Optical-to-Electrical-to-Optical (O/E/O) conversion and retransmission by a hub node.  Even so, on average the $R=2$ and $R=3$ redundant quorums cycle solutions would require infrequent additional steps considering the missing pairs are less than 1\% on average.

\subsubsection{Fault-tolerant operational analysis} \label{chpt:COMM:sub:RedundantSingle:Fault-Tolerance}

Using the generalized $R$ quorum redundancy rather than cycle pairs can save significant resources; however, this can come at a significant determent to fault tolerance.

To model the fault, we simulated the failure of each used edge, $(e_i,e_j) \in E$, in 100 node mappings of each network model, $G=(V,E)$.  The edges not used in a particular mapping are ignored to prevent biasing the results with zero missing pairs data points. We then examine the network's ability to serve all potential point-to-point requests by counting pairs of nodes that would be able to communicate and conversely those pairs that are unable to communicate.  The results are then reported as fault coverage.  To calculate fault coverage in this scenario, we calculated the mean connected pairs divided by the total pairs.
\begin{align}
1-\frac{Mean\: Missing\: Pairs}{Total\: Pairs}
\end{align}
100\% would be perfect coverage, whereas 0\% would be no fault coverage at all.

		

\begin{figure}[t]
	\centering
	\includegraphics[width=2.5in]{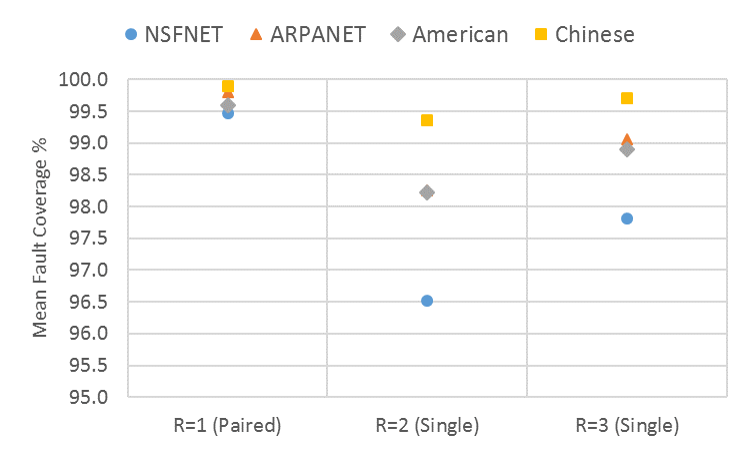}
	\caption{Percent mean fault coverage of the single cycle, redundant quorum solution experiencing a single link fault.}
	\label{fig:Single:OneFault}
\end{figure}

Our simulation results (Fig. \ref{fig:Single:OneFault}) shows the redundant quorum-based cycle technique had 96.52 - 99.36\%  and 97.81 - 99.71\% fault coverages, $R=2$ and $R=3$ respectively, in the four networks tested.  We compare the state-of-art paired cycle approach with our quorum redundant technique with single cycles that uses significantly fewer resources.  With single link failures, the paired cycles had a mean missing communication pair rate of less than 3 pairs or less than 0.53\% across all networks (95\% CI). Hence the $R=1$ (Paired) column shows mean fault coverage percentages is greater than 99.47\% for all four networks.  The redundant quorum cycles technique, $R=2$ and $R=3$ (Single), could not reach that level of coverage, but did achieve a mean fault coverage rate (95\% CI) of greater than 96.52 and 97.81\%, respectively, across all networks.

While neither single cycle $R=2$ or $R=3$ could achieve the same level of fault coverage as the paired cycle solution, they did have missing pair rates better than 3.48 and 2.29\%, respectively, while achieving significant resource savings.  In networks where an additional approximately 40\% of resources could better be utilized for communication rather than redundancy, the trade off of missing a relatively small percentage of communications during fault conditions may be considered tolerable.

\section{Improving Single Cycle Routing Based on Redundant Cyclic Quorums} \label{sec:improvingHeuristic}

In Section \ref{chpt:COMM:sub:RedundantSingle:Fault-Free}, we had an interesting observation that although the cyclic quorums set method was proven to guarantee that all of the node pairs exist, the implementation that utilized the quorums still had the limitation of unidirectional light-trail cycles requiring one optical shutter in the \textit{off} state per cycle.  This led to missing pairs even in fault-free states of the network and likely contributed to at least some of missing pairs during fault scenarios.

When we used paired cycles, the routing order within a cycle did not matter.  In any given pair of cycles, there was one in the forward direction and the other in the backwards.  For example, if the forward cycle route formed the $0\rightarrow 1$ communication route, then the backward cycle route would form the $0\leftarrow 1$ communication route.  

With a single, unidirectional cycle we were taking for granted that redundant cyclic quorums were guaranteeing two pairs of all node pairs.  However, the route direction of the pairs is not enforced.  Hence, without the backwards cycle from paired cycle, it turned out that the single cycle could still result in two forward $0\rightarrow 1$ paths in the network's cycle route implementation rather than one of each ($0\rightarrow 1$ and $0\leftarrow 1$ communication routes).

Looking closer at the source of the problem, i.e., the unidirectional optical light-trail cycles, we developed an alternative heuristic solution to improve the performance.  The cycles were routed using the ECBRA heuristic\cite{dlastine2011ECBRA} and the redundant cyclic quorums for $R=2$ and $R=3$.  Once routed, we treated the first node in the corresponding cyclic quorum to be the hub node and then formed the cycle in the order provided by the ECBRA heuristic.  By controlling the order to either be forward or backwards on a particular unidirectional optical light-trail cycle, we can influence which node pairs are formed.

\subsection{Greedy cycle direction based on missing pairs} \label{chpt:COMM:sub:GreedyAlg}

There are $\left|V\right|=N$ redundant cyclic quorums for a given network $G=(V,E)$.  Each of these quorums has a corresponding cycle route and each route is either in the forward or backward direction.  This results in $O\left(2^N\right)$ possible combinations of cycle directions for a network.

Next, we describe our greedy algorithm to determine cycle direction and evaluate the improvement in network performance.  Ultimately, by controlling the forwards or backwards direction of a cycle, we want to decrease (and eliminate) the number of missing pairs under fault-free conditions.  While doing this, the network's fault tolerance is also anticipated to increase.

Algorithm \ref{alg_initial_cycle_direction} greedily chooses each cycle's initial direction.  All forward and backward node pairs need to be formed in the network.  We keep track of how many of each pair have been formed in variable $PC$ on Line \ref{alg_ICD_PC}.  We iterate through all of the routed cycles, one for each redundant quorum (Line \ref{alg_ICD_forall}).  The cycle's direction is chosen by which direction will eliminate more missing pairs from $PC$.  Then all of that direction's pairs are added to $PC$ (Line \ref{alg_ICD_forward} or Line \ref{alg_ICD_backward}).  The next cycle's direction is chosen the same way until all cycles have been assigned an initial direction.  

\begin{algorithm}
	\caption{Initial Cycle Direction($Cycles$,$V$)}\label{alg_initial_cycle_direction}
	\begin{algorithmic}[1]
		\STATE Pair Count $PC\left[e_i\right]\left[e_j\right]\gets 0, \:\forall e_i,e_j\in V$ \label{alg_ICD_PC}
		\FORALL {$c \in Cycles$}\label{alg_ICD_forall}
		\STATE Count number of new pairs added to $PC$ if Cycle $c.direction = Forward$
		\STATE Count number of new pairs added to $PC$ if Cycle $c.direction = Backward$
		\IF {Forward count $\ge$ Backward count}
		\STATE $c.direction \gets Forward$
		\STATE Increment $PC\left[e_i\right]\left[e_j\right]$ for each forward direction pair from Cycle c\label{alg_ICD_forward}
		\ELSE
		\STATE $c.direction \gets Backward$
		\STATE Increment $PC\left[e_i\right]\left[e_j\right]$ for each backward direction pair from Cycle c\label{alg_ICD_backward}
		\ENDIF
		\ENDFOR
	\end{algorithmic}
\end{algorithm}

There are $\left|V\right|=N$ redundant cyclic quorums for a given network.  Each of these quorums has a corresponding cycle route in $Cycles$ causing Algorithm \ref{alg_initial_cycle_direction}'s For-loop on Line \ref{alg_ICD_forall} to execute $O\left(N\right)$ times.  Both forward and backward new pair counts, as well as, the final incrementing of $PC$ after a direction is chosen requires enumerating all possible pairs in a particular cycle.  From Equation \ref{eq:Rsize}, each quorum (i.e., cycle) will be of size approximately $O\left(\sqrt{RN}\right)$.  Forming all pairs is $\binom{N}{2}=O\left(N^2\right)$ operation, so $O\left(\sqrt{RN}^2\right)=O\left(RN\right)$.  All combined this results in an $O\left(N*RN\right)=O\left(RN^2\right)$ runtime to find the initial cycle direction.

The order of cycle iteration has an impact on the final direction of all cycles.  It is possible that a cycle processed later may add pairs to $PC$ that a previous cycle had already contributed.  This opens up the possibility that a cycle processed earlier may be able to change to a more favorable direction and further reduce the number of missing pairs.  Because of this, we have a second greedy heuristic, Algorithm \ref{alg_GreedyUpdate_cycle_direction}.  

\begin{algorithm}
	\caption{Greedy Update Cycle Direction($Cycles$,$V$,$PC$)}\label{alg_GreedyUpdate_cycle_direction}
	\begin{algorithmic}[1]
		\STATE $Changed \gets True$ \label{alg_GU_changedStart}
		\WHILE {$Changed \ne False$}
		\STATE $Changed \gets False$ \label{alg_GU_changedEnd}
		\FORALL {$c \in Cycles$} \label{alg_GU_forall}
		\STATE Decrement $PC\left[e_i\right]\left[e_j\right]$ for each pair from Cycle c
		\STATE Count number of new pairs added to $PC$ if Cycle $c.direction = Forward$
		\STATE Count number of new pairs added to $PC$ if Cycle $c.direction = Backward$
		\IF {Forward count $>$ Backward count}
		\STATE Increment $PC\left[e_i\right]\left[e_j\right]$ for each forward direction pair from Cycle c \label{alg_GU_forward}
		\IF {$c.direction \ne Forward$}
		\STATE $c.direction \gets Forward$
		\STATE $Changed \gets True$
		\ENDIF
		\ELSIF {Backward count $>$ Forward count}
		\STATE Increment $PC\left[e_i\right]\left[e_j\right]$ for each backward direction pair from Cycle c \label{alg_GU_backward}
		\IF {$c.direction \ne Backward$}
		\STATE $c.direction \gets Backward$
		\STATE $Changed \gets True$
		\ENDIF
		\ELSE \label{alg_GU_equal}
		\STATE Increment $PC\left[e_i\right]\left[e_j\right]$ for each pair from Cycle c
		\ENDIF
		\ENDFOR
		\ENDWHILE
	\end{algorithmic}
\end{algorithm}

This algorithm modifies the direction of a single cycle at a time to improve the missing pairs count.  It continues modifying cycle directions until no further single cycle direction flips result in improvements being made.  Although this appears to be an infinite loop on the surface, its runtime is considerably faster than evaluating all $O\left(2^N\right)$ possible combinations of cycle directions for a network.

Algorithm \ref{alg_GreedyUpdate_cycle_direction} uses a variable $Changed$ to determine if it should continue searching for a better combination of cycle directions (Lines \ref{alg_GU_changedStart}-\ref{alg_GU_changedEnd}).  Line \ref{alg_GU_forall} iterates through all of the routed cycles.  The pair count variable $PC$ from the cycle direction initialization is utilized once again to account for the number of node pairs formed and also for how many pairs are missing.  The cycle's direction is then chosen by whichever direction will eliminate more missing pairs from $PC$.  Then all of that direction's pairs are added to $PC$ (Line \ref{alg_GU_forward} or Line \ref{alg_GU_backward}).  If this selection causes the cycle's direction to change, then the $Changed$ variable is set to $True$ so the While-loop knows to continue searching.  If the backward and forward missing pair counts are the same, then the direction defaults to the direction the cycle is currently (Line \ref{alg_GU_equal}).  This process continues until an entire iteration of $Cycles$ results in no direction changes.

The outer While-loop continues until there are no direction changes.  A direction change will only occur if it results in an improvement in the number of missing pairs.  There are $\left|V\right|=N$ nodes and forming all pairs is $\binom{N}{2}=O\left(N^2\right)$ operation.  So, even if every iteration of the While-loop only improved the number of missing pairs by 1, it would still only loop $O\left(N^2\right)$ times.  The For-loop on Line \ref{alg_GU_forall} will execute $O\left(N\right)$ times as it iterates over all cycles.  Both forward and backward new pair counts, as well as, the initial decrement and final incrementing of $PC$ after a direction is chosen requires enumerating all possible pairs in a particular cycle, hence $O\left(RN\right)$, same as was calculated for Algorithm \ref{alg_initial_cycle_direction}.  All combined, this results in an $O\left(N^2*N*RN\right)=O\left(RN^4\right)$ runtime to greedily update the cycle direction.  

A 4th power is certainly not desired for most algorithms; but compared to $O\left(2^N\right)$, it scales well.  For $N$ greater than 16 this greedy approach is better than the brute force approach.  And certainly for $N \le 16$, the outer While-loop in Algorithm \ref{alg_GreedyUpdate_cycle_direction} will typically not behave in the worst case having to execute $O\left(N^2\right)$ times.  In fact, given results from Section \ref{chpt:COMM:sub:RedundantSingle:Fault-Free}, on average less than 1\% of the pairs were missing.  Therefore, we can expect that the runtime behavior of the While-loop to be quite small for network inputs similar to those in our analysis.

\subsection{Greedy missing pairs heuristic results}

This section uses a similar experiment setup as Section \ref{sec:singleCycleAnalysis}.  We used the same four common networks (Fig. \ref{fig:Networks}) and an implementation of the ECBRA heuristic\cite{dlastine2011ECBRA} to perform the cycle routing.  We also are using just a single cycle based on $R$ redundant quorums sets.

\subsubsection{Fault-free operational analysis} 

Section \ref{chpt:COMM:sub:RedundantSingle:Fault-Free} showed significant resource usage reductions freeing up (wavelength) resource availability within the network and adding the potential for lower capital costs in terms of physical transmitters and receivers needed.  The challenge, though, was this came at a cost.  There were missing communication node pairs even in fault-free operation of the network.

Table \ref{tbl:greedy:fault-free} shows the significant improvements that our greedy heuristic had on this issue.  An average reduction of greater than 94\% of missing pairs for single cycle routing based on $R=2$ redundant cyclic quorums.  And a complete elimination of missing pairs for $R=3$ redundancy.

\begin{table*}[t]
	\centering
	\caption{Comparing fault-free operation mean percent missing node pairs (95\% CI) by single cycles using redundant quorums and our greedy cycle direction heuristic}
	\label{tbl:greedy:fault-free}
	\begin{tabular}{@{}lccccccccc@{}}
		\toprule
		& 
		$\mathbf{R=1}$    & 
		\multicolumn{4}{c}{$\mathbf{R=2}$ \textbf{(Single) (\%)}} &
		\multicolumn{4}{c}{$\mathbf{R=3}$ \textbf{(Single) (\%)}} \\ \cline{3-10} 
		
		\textbf{Network} & 
		\textbf{(Paired) (\%)}             &
		\textbf{Forward}             &
		\textbf{Random}    & 
		\textbf{Greedy}             &
		\textbf{Reduction}                 &
		\textbf{Forward}             &
		\textbf{Random}    & 
		\textbf{Greedy}             &
		\textbf{Reduction}    \\
		\midrule
		
		NSFNET   & 0.00 $\pm$ 0.00 & 0.95 $\pm$ 0.15 & 0.85 $\pm$ 0.12 & 0.02 $\pm$ 0.02 & -98.27 & 0.04 $\pm$ 0.03 & 0.85 $\pm$ 0.12 & 0.00 $\pm$ 0.00 & -100 \\
		ARPANET  & 0.00 $\pm$ 0.00 & 0.36 $\pm$ 0.07 & 0.31 $\pm$ 0.06 & 0.01 $\pm$ 0.01 & -97.78 & 0.13 $\pm$ 0.04 & 0.31 $\pm$ 0.06 & 0.00 $\pm$ 0.00 & -100 \\
		American & 0.00 $\pm$ 0.00 & 0.49 $\pm$ 0.07 & 0.52 $\pm$ 0.07 & 0.02 $\pm$ 0.01 & -95.56 & 0.21 $\pm$ 0.04 & 0.52 $\pm$ 0.07 & 0.00 $\pm$ 0.00 & -100 \\
		Chinese  & 0.00 $\pm$ 0.00 & 0.27 $\pm$ 0.03 & 0.26 $\pm$ 0.02 & 0.01 $\pm$ 0.00 & -94.98 & 0.09 $\pm$ 0.01 & 0.26 $\pm$ 0.02 & 0.00 $\pm$ 0.00 & -100 \\ \bottomrule
	\end{tabular}
\end{table*}

Column 2 in Table \ref{tbl:greedy:fault-free} is the prior art paired cycle solution.  This requires at minimum 38\% more link resources on average, but no communication node pairs are missing.  Originally in Section \ref{sec:singleCycleAnalysis}, the cycle routing directly from the ECBRA heuristic was used.  This produced the percent missing pairs.

The emphasis on the greedy heuristic was that chosen direction of the cycle matters.  To emphasize this point and to confirm that our approach is actually doing something intelligent, we also included a random cycle direction algorithm (Column 4 and 8).  There we see the random direction on average performs similar (and at times worse) than simply using the forward cycle directions directly from the ECBRA heuristic.  Columns 5 and 6 (9 and 10 respectively) show the impact from our greedy heuristic.  Nearly all of the missing pairs are removed on average from the $R=2$ redundant quorum cycles and all of the missing pairs are eliminated from the $R=3$ redundant quorum cycles.  This makes the $R$ redundant quorum, single cycle solution with greedy cycle direction heuristic a significant improvement over prior art's paired cycle approach.  All communication pairs are formed in the network and done so with significantly fewer resources.

\subsubsection{Fault-tolerant operational analysis}

Section \ref{chpt:COMM:sub:RedundantSingle:Fault-Tolerance} illustrated a trade off between the significant reduction to resource usage and maintaining the same level of fault tolerance that prior art offered.  Using only single cycles had a small impact on fault tolerance of the optical network, which for some networks and applications may still be acceptable given the improved resource usage.  With the greedy heuristic addressing missing communication pair issues in the prior section, this section looks at the improvements to fault tolerance.

To model the fault(s), we simulate the failure of used edge(s), $(e_i,e_j) \in E$, in the 100 node mappings of each network model, $G=(V,E)$.  The edges not used are ignored. We then examine the network's ability to serve all potential point-to-point requests.  The results are then reported as fault coverage, total pairs able to communicate as a percentage of total point-to-point pairs.  100\% would be perfect fault coverage, whereas 0\% is no fault coverage at all.

\begin{table*}[t]
	\centering
	\caption{Comparing percent mean fault coverage (95\% CI) of the single cycle, redundant quorum solution with our greedy cycle direction heuristic experiencing a single link fault.}
	\label{tbl:Greedy:OneFault}
	\begin{tabular}{@{}lccccccc@{}}
		\toprule
		& 
		$\mathbf{R=1}$    & 
		\multicolumn{3}{c}{$\mathbf{R=2}$ \textbf{(Single) (\%)}} & 
		\multicolumn{3}{c}{$\mathbf{R=3}$ \textbf{(Single) (\%)}}\\ \cline{3-8} 
		
		\textbf{Network} & 
		\textbf{(Paired) (\%)}             &
		\textbf{Forward}             &
		\textbf{Random}    & 
		\textbf{Greedy}    &
		\textbf{Forward}             &
		\textbf{Random}    & 
		\textbf{Greedy}    \\
		\midrule
		
		NSFNET   & 99.47 $\pm$ 0.04 & 96.52 $\pm$ 0.09 & 96.84 $\pm$ 0.08 & 97.92 $\pm$ 0.07 & 97.81 $\pm$ 0.09 & 96.84 $\pm$ 0.08 & 98.59 $\pm$ 0.06 \\
		ARPANET  & 99.81 $\pm$ 0.01 & 98.25 $\pm$ 0.05 & 98.28 $\pm$ 0.05 & 98.79 $\pm$ 0.04 & 99.05 $\pm$ 0.04 & 98.28 $\pm$ 0.05 & 99.42 $\pm$ 0.03 \\
		American & 99.60 $\pm$ 0.02 & 98.23 $\pm$ 0.04 & 98.21 $\pm$ 0.04 & 98.92 $\pm$ 0.03 & 98.90 $\pm$ 0.03 & 98.21 $\pm$ 0.04 & 99.34 $\pm$ 0.02 \\
		Chinese  & 99.90 $\pm$ 0.00 & 99.36 $\pm$ 0.01 & 99.38 $\pm$ 0.01 & 99.67 $\pm$ 0.01 & 99.71 $\pm$ 0.00 & 99.38 $\pm$ 0.01 & 99.84 $\pm$ 0.00 \\ \bottomrule
	\end{tabular}
\end{table*}

Our single fault simulation results (Table \ref{tbl:Greedy:OneFault}) shows the redundant quorum-based cycle technique with our greedy cycle direction heuristic had 97.92 - 99.67\%  and 98.59 - 99.84\% fault coverages, $R=2$ and $R=3$ respectively, in the four networks tested.  This was due to a 31.16 - 48.74\% and 35.48 - 44.85\%, $R=2$ and $R=3$ respectively, improvement over the number of missing pairs when only using forward cycle directions (i.e., Section \ref{chpt:COMM:sub:RedundantSingle:Fault-Tolerance}).  Again, we also test the performance against random cycle direction choices (Columns 4 and 7).  This illustrates that the greedy heuristic, while it was designed to eliminate missing pairs for fault-free network operation, still provides an intelligent improvement for fault conditions as well.

With single link failures, the paired cycles had a mean missing communication pair rate of less than 3 pairs or less than 0.53\% across all networks (95\% CI). Hence the $R=1$ (Paired) column shows mean fault coverage percentages is greater than 99.47\% for all four networks.  The redundant quorum cycles technique with our greedy heuristic, $R=2$ and $R=3$ (Single), reduced the number of missing pairs by greater than 30\% when compared to without the heuristic (i.e., all forward cycles).  Even with the improvement though, the fault tolerance still could not reach the level of coverage that the prior art's paired cycles.  The Table \ref{tbl:Greedy:OneFault} does show a competitive mean fault coverage rate (95\% CI) of greater than 97.92 and 98.59\%, respectively, across all networks when our greedy cycle direction heuristic was used with the single quorum cycle solution.

\section{Conclusion}
The $R$ redundant quorums significantly reduced resource usage and maintained high fault tolerance capabilities.  The single cycle technique used significantly fewer resources (42.91 - 47.19\% and 38.85 - 42.39\% fewer, respectively), while at the same time maintained a high degree of fault tolerance with 96.52 - 99.36\% and 97.81 - 99.71\% fault coverage, respectively, on the single fault simulation.  

With the paired cycles in prior art, the direction of communication pairs were not a consideration because both directions were always present.  When we switched to single cycles, the resource usage fell dramatically (greater than 38\%), and there was no longer the guarantee that both directions of a communication pair existed.  To address this limitation, we developed a greedy algorithm to determine whether a cycle should be routed in the forward or backward direction in order to eliminate the most missing pairs and achieve higher fault tolerance.  Missing pairs were reduced by over 94\% for $R=2$ redundant quorum single cycles and eliminated completely for $R=3$.  Fault tolerance improved as well.  Missing pairs during single link failures decreased by over 30\%, increasing fault tolerance to 97.92\% - 99.67\% and 98.59\% - 99.84\%, respectively.

\section*{Acknowledgment}

Research funded in part by NSF Graduate Research Fellowship Program, IBM Ph.D. Fellowship Program, Symbi GK-12 and Trinect Fellowships at Iowa State University, and the Jerry R. Junkins Endowment at Iowa State University.  The research reported in this paper is partially supported by the HPC@ISU equipment at Iowa State University, some of which has been purchased through funding provided by NSF under MRI grant number CNS 1229081 and CRI grant number 1205413.  Any opinions, findings, and conclusions or recommendations expressed in this material are those of the author(s) and do not necessarily reflect the views of the funding agencies.


\bibliographystyle{IEEEtran}
\bibliography{../latex_bibs/mybib}

\end{document}